\documentclass[aps,prd,reprint,amsmath,amssymb,superscriptaddress,nofootinbib]{revtex4-1}

\usepackage{graphicx}% Include figure files
\usepackage{bm}% bold math
\usepackage{hyperref}
\usepackage{color}

\newcommand{\mnras}{MNRAS}
\newcommand{\aap}{A\&A}

\newcommand{\apjs}{ApJS}

\newcommand{\jhep}{JHEP}

\newcommand{\jcap}{JCAP}
\newcommand{\pasp}{PASP}
\newcommand{\pnas}{PNAS}
\newcommand{\epjc}{EPJC}

\newcommand{\planck}{{\it Planck}}
\newcommand{\plancks}{{\it Planck{\rm 's}}}
\newcommand{\lcdm}{$\Lambda$CDM}
\newcommand{\pars}{\ensuremath{{\bm{\theta}}}}
\newcommand{\msn}{\ensuremath{{\bm{m}}}}
\newcommand{\msnobs}{\ensuremath{{\hat{\bm{m}}}}}
\newcommand{\covsn}{\ensuremath{{\bm{\Sigma}_{\rm s}}}}
\newcommand{\abao}{\ensuremath{{\bm{\alpha}}}}
\newcommand{\abaoobs}{\ensuremath{{\hat{\bm{\alpha}}}}}
\newcommand{\covbao}{\ensuremath{{\bm{\Sigma}_{\rm b}}}}
\newcommand{\kmsmpc}{\ensuremath{{\rm km\,s^{-1}\,Mpc^{-1}}}}
\newcommand{\hubble}{\ensuremath{H_0}}
\newcommand{\hz}{\ensuremath{H(z)}}

\newcommand{\hubbleobs}{\ensuremath{\hat{H}_0}}
\newcommand{\decel}{\ensuremath{q_0}}
\newcommand{\jerk}{\ensuremath{j_0}}
\newcommand{\rdrag}{\ensuremath{r_{\rm d}}}

\newcommand{\dm}{\ensuremath{d_{\rm M}(z)}}

\newcommand{\dl}{\ensuremath{d_{\rm L}(z)}}
\newcommand{\vpec}{\ensuremath{v^{\rm p}}}
\newcommand{\vpecobs}{\ensuremath{\hat{v}^{\rm p}}}
\newcommand{\zobs}{\ensuremath{\hat{z}}}
\newcommand{\dobs}{\ensuremath{\hat{d}}}
\newcommand{\prob}{\ensuremath{{\rm Pr}}}
\newcommand{\nbns}{\ensuremath{n}}
\newcommand{\nk}{\ensuremath{m}}
\newcommand{\RBNS}{\ensuremath{3000\,{\rm Gpc^{-3}\,yr^{-1}}}}
\newcommand{\NBNS}{\ensuremath{51}}

\begin{document}
\title{Prospects for resolving the Hubble constant tension with standard sirens}
\author{Stephen M. Feeney}
\email{sfeeney@flatironinstitute.org}
\affiliation{Center for Computational Astrophysics, Flatiron Institute, 162 5th Avenue, New York, NY 10010, USA}
\author{Hiranya V. Peiris}
\affiliation{Department of Physics \& Astronomy, University College London, Gower Street, London WC1E 6BT, UK}
\affiliation{Oskar Klein Centre for Cosmoparticle Physics, Stockholm University, AlbaNova, Stockholm SE-106 91, Sweden}
\author{Andrew R. Williamson}
\affiliation{Institute of Mathematics, Astrophysics and Particle Physics, Radboud University, Heyendaalseweg 135, 6525 AJ Nijmegen, The Netherlands}
\author{\\Samaya M. Nissanke}
\affiliation{Institute of Mathematics, Astrophysics and Particle Physics, Radboud University, Heyendaalseweg 135, 6525 AJ Nijmegen, The Netherlands}
\author{Daniel J. Mortlock}
\affiliation{Astrophysics Group, Imperial College London, Blackett Laboratory, Prince Consort Road, London SW7 2AZ, UK}
\affiliation{Department of Mathematics, Imperial College London, London SW7 2AZ, UK}
\affiliation{Department of Astronomy, Stockholm University, AlbaNova, SE-10691 Stockholm, Sweden}
\author{Justin Alsing}
\affiliation{Center for Computational Astrophysics, Flatiron Institute, 162 5th Avenue, New York, NY 10010, USA}
\author{Dan Scolnic}
\affiliation{Kavli Institute for Cosmological Physics, The University of Chicago, Chicago, IL 60637, USA}

%%%%%%%%%%%%%%%%%%%%%%%%%%%%%%%%%%%%%%%%%%%%%%%%%%
%%%%%%%%%%%%%%%%%%%%%%%%%%%%%%%%%%%%%%%%%%%%%%%%%%

\begin{abstract}
The Hubble constant (\hubble) estimated from the local Cepheid-supernova (SN) distance ladder is in 3-$\sigma$ tension with the value extrapolated from cosmic microwave background (CMB) data assuming the standard cosmological model. Whether this tension represents new physics or systematic effects is the subject of intense debate. Here, we investigate how new, independent \hubble\ estimates can arbitrate this tension, assessing whether the measurements are consistent with being derived from the same model using the posterior predictive distribution (PPD). We show that, with existing data, the inverse distance ladder formed from BOSS baryon acoustic oscillation measurements and the Pantheon SN sample yields an \hubble\ posterior near-identical to the \planck\ CMB measurement. The observed local distance ladder value is a very unlikely draw from the resulting PPD. Turning to the future, we find that a sample of $\sim50$ binary neutron star ``standard sirens'' (detectable within the next decade) will be able to adjudicate between the local and CMB estimates.
\end{abstract}

%%%%%%%%%%%%%%%%%%%%%%%%%%%%%%%%%%%%%%%%%%%%%%%%%%
%%%%%%%%%%%%%%%%%%%%%%%%%%%%%%%%%%%%%%%%%%%%%%%%%%

\maketitle

\section{Introduction}
\label{sec:intro}

The Hubble constant (\hubble)---the current expansion rate of the Universe~\cite{Hubble:1929}---is one of few cosmological parameters that can be estimated locally, using a minimal physical model. Such measurements are invaluable in breaking degeneracies between \hubble\ and other cosmological parameters (e.g., the spatial curvature of the Universe or number/mass of neutrinos). A plethora of methods exist to estimate \hubble, using Cepheid variables, red-giant stars, SNe, gravitational lenses, galaxies, the CMB and neutron-star mergers~\cite[most recently][]{Aubourg_etal:2015,Planck_XIII:2016,Riess_etal:2016,Jang_etal:2017,Bonvin_etal:2017,Addison_etal:2017,Henning_etal:2018,Abbott_etal:2017a,DES_H_0:2017,Vega-Ferrero_etal:2017,Gomez-Valent_etal:2018}. The best cosmology-independent constraints come from the SH0ES Cepheid-SN distance ladder~\cite{Riess_etal:2016}; the tightest constraints come from the \planck\ CMB data, assuming a standard \lcdm\ cosmology~\cite{Planck_XIII:2016}. These estimates are discrepant at the 3-$\sigma$ level, suggesting the possibility that the measurements contain unmodeled systematics or that \lcdm\ is not the true cosmology~\cite{Feeney_etal:2017}.

Numerous attempts have been made to reconcile the two results through new physics~\cite{Wyman_etal:2014,*Pourtsidou_etal:2016,*Di_Valentino_etal:2016,*Huang_etal:2016,*Bernal_etal:2016,*Ko_etal:2016,*Karwal_etal:2016,*Kumar_etal:2016,*Santos_etal:2017,*Prilepina_etal:2017,*Zhao_etal:2017,*Zhao_M_etal:2017,*Solar_etal:2017,*Di_Valentino_etal:2017a,*Di_Valentino_etal:2017b} or improved astrophysical, experimental and statistical modeling~\cite{Efstathiou:2014,*Spergel_etal:2015,*Rigault_etal:2015,*Jones_etal:2015,*Addison_etal:2016,*Planck_Int_XLVI:2016,*Cardona_etal:2016,*Zhang_etal:2017,*Wu_Huterer:2017,Feeney_etal:2017,*Follin_Knox:2017,*Dhawan_etal:2017}, yielding no compelling explanation. Here, we look to the inverse distance ladder and gravitational wave (GW) standard sirens~\cite{Schutz:1986} to provide the independent information needed arbitrate this tension, which we frame in a new, intuitive way using the posterior predictive distribution (PPD). Unlike existing tension metrics based on the ``$n$-$\sigma$'' discrepancy~\cite[{e.g.,}][]{Riess_etal:2016}, Kullback-Leibler divergence~\cite[{e.g.,}][]{Seehars_etal:2016,Wang_etal:2017} or Bayesian evidence ratio~\cite[{e.g.,}][]{Marshall_etal:2006,Feeney_etal:2017}, the PPD is simple to interpret and cheap to calculate for non-Gaussian distributions, and does not require the specification of a (potentially arbitrary) alternative model.

%%%%%%%%%%%%%%%%%%%%%%%%%%%%%%%%%%%%%%%%%%%%%%%%%%
%%%%%%%%%%%%%%%%%%%%%%%%%%%%%%%%%%%%%%%%%%%%%%%%%%

\section{Quantifying Tension}
\label{sec:methods}

{\bf Inverse Distance Ladder.} We first demonstrate the PPD's utility as a tension metric using the inverse distance ladder constructed from existing baryon acoustic oscillation (BAO) and Type Ia SN observations~\cite{Aubourg_etal:2015}. Galaxy redshift surveys measure the BAO scale parallel and perpendicular to the line of sight, $\alpha^\parallel$ and $\alpha^\perp$. These are linked to the sound horizon at radiation drag, \rdrag, the Hubble parameter $H(z)$ at the redshift $z$ of the observations, and the transverse comoving distance~\cite{Hogg:1999} $\dm = \int_0^z c \, dz^\prime / H(z^\prime)$ (assuming a flat universe), by~\cite{Padmanabhan_etal:2008,Alam_etal:2017}
\begin{equation}
\alpha^\parallel = \frac{ \left[\hz \, \rdrag \right]_{\rm fid} }{ \hz \, \rdrag } \,\,\,\,{\rm and}\,\,\,\,\, \alpha^\perp = \frac{ \dm }{ \rdrag } \left[ \frac{ \rdrag}{ \dm } \right]_{\rm fid},
\label{eq:x_bao}
\end{equation}
where the comparison is to a fiducial cosmology. Given a CMB measurement of \rdrag, a BAO survey at redshift $z$ therefore constrains both \hz\ and $\dm$.

By adopting a model for \hz, the BAO measurements can be extrapolated to redshift zero and hence converted to estimates of \hubble; however, additional data are required to constrain flexible models. Modern SN surveys are ideal for this task, providing $\mathcal{O}(10^3)$ relative distance measurements over the relevant redshift range.
The apparent magnitude $m$ of a SN of absolute magnitude $M$ probes the luminosity distance, $\dl = (1+z) \, \dm$, via
\begin{equation}
m = 5 \log_{10} \left( \frac{\dl}{\rm pc} \right) + M - 5.
\label{eq:mu}
\end{equation}
The absolute distance scale of a pure-SN dataset is completely degenerate with the unknown value of $M$, but combining with BAO data (transverse measurements in particular) breaks this degeneracy, allowing precise determination of the distance-redshift relation well into the linear regime. The resulting inverse distance ladder prefers \cite{Aubourg_etal:2015,Cuesta_etal:2015,Addison_etal:2017} values of \hubble\ in close agreement with that of the \planck\ flat \lcdm\ analysis, and is thus in tension with the SH0ES Cepheid distance ladder estimate. The recent release of the Pantheon SNe sample~\cite{Scolnic_etal:2017}---with 50\% greater statistical power than the previous gold standard~\cite{Betoule_etal:2014} and a full recalibration of all subsamples used~\cite{Scolnic_etal:2015}---strongly motivates revisiting this analysis.

In order not to restrict ourselves to a particular physical model, we assume only that the expansion is smooth, adopting the third-order Taylor expansion of the luminosity distance (used by SH0ES):
\begin{equation}
\dl = \frac{c\,z}{\hubble} \left[ 1 + \frac{z}{2}(1 - \decel) - \frac{z^2}{6}\left(1 - \decel - 3\decel^2 + \jerk\right) \right],
\label{eq:d_l_exp}
\end{equation}
where \decel\ and \jerk\ are the deceleration and jerk parameters.\footnote{While the Taylor expansion is designed for $z < 1$, it performs well in our redshift range: modeling the expansion history as a Gaussian Process (following Ref.~\cite{Shafieloo_etal:2012}) instead yields near-identical $\hubble$ posteriors.} Our inverse distance ladder therefore depends on only five parameters, $\pars= \{\hubble, \decel, \jerk, \rdrag, M \}$. Given $n_s$ observed SN apparent magnitudes $\msnobs$ and $n_b$ BAO observations $\abaoobs$, the joint posterior of these parameters is
\begin{equation}
\prob(\pars|\msnobs,\abaoobs,I) \propto \prob(\pars|I) \, {\rm N}(\msnobs; \msn, \covsn) \, {\rm N}(\abaoobs; \abao, \covbao),
\label{eq:idl_posterior}
\end{equation}
where the theoretical SN magnitudes \msn\ and BAO measurements \abao\ are given by Eqs.~\ref{eq:x_bao} to~\ref{eq:d_l_exp}, ${\rm N}(\bm{x};\bm{\mu},\bm{\Sigma})$ is a multivariate normal distribution with mean $\bm{\mu}$ and covariance $\bm{\Sigma}$, and $\covsn$ and $\covbao$ are the SN and BAO covariance matrices. We adopt uniform priors on all parameters apart from \rdrag, for which we assume a Gaussian prior derived from CMB observations. We sample the joint posterior distribution (Eq.~\ref{eq:idl_posterior}) using {\tt emcee}~\cite{Foreman-Mackey_etal:2013}.

\begin{figure}
\includegraphics[width=8.5cm]{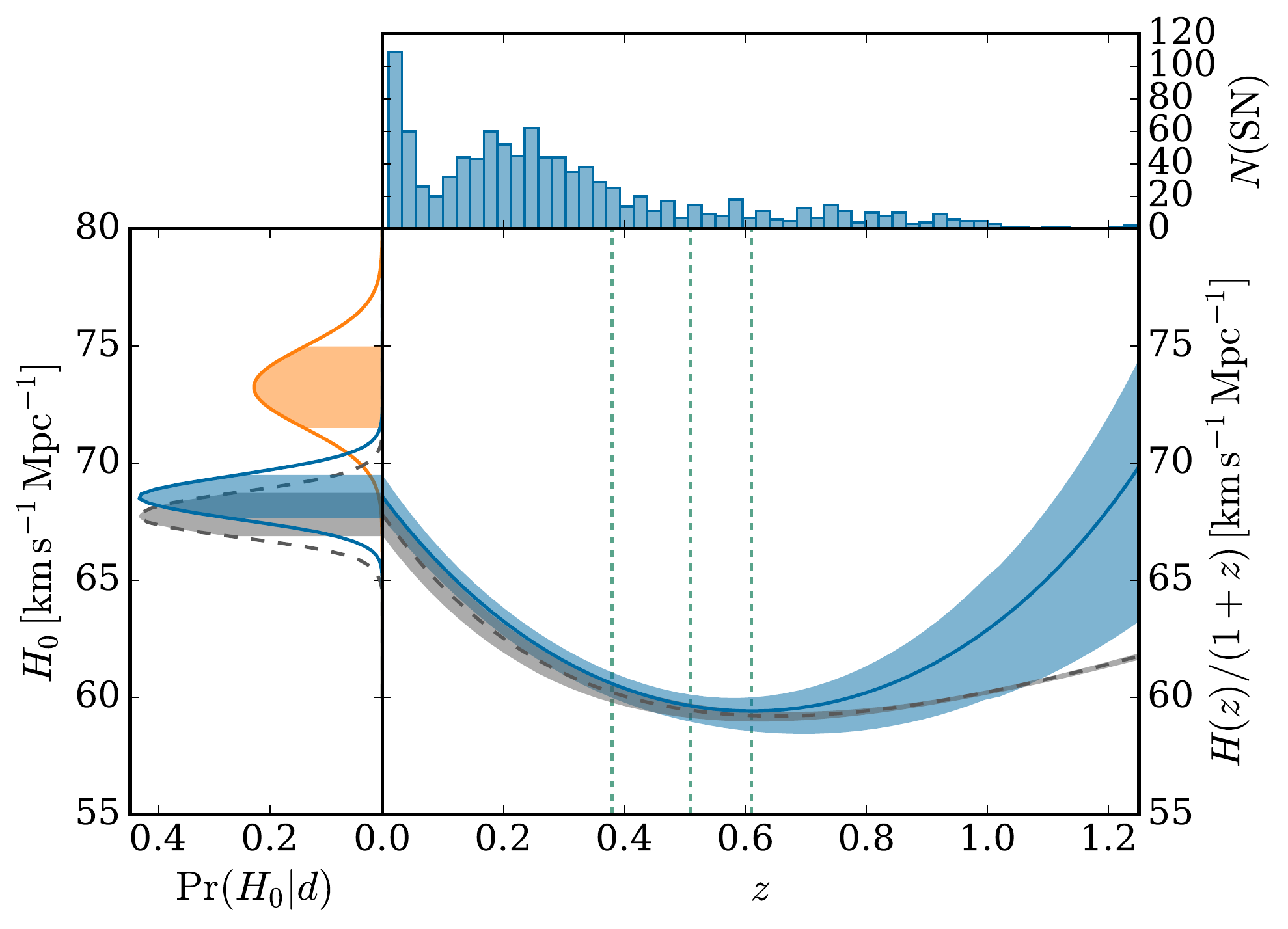}
\caption{Main panel: expansion history for BOSS BAO, Pantheon SNe and \planck\ \rdrag\ assuming smooth expansion and early-time physics only (blue), and for \planck\ assuming \lcdm\ (grey). BAO redshifts are shown as short-dashed lines. Left panel: corresponding \hubble\ posteriors and Cepheid distance ladder measurement (orange). Top panel: redshift distribution of Pantheon SNe.}
\label{fig:idl_h_z}
\end{figure}

Combining the BOSS DR12 BAO measurements~\cite{Alam_etal:2017} with the Pantheon SN sample~\cite{Scolnic_etal:2017} and \plancks\ ``TT+lowP+lensing'' \rdrag\ posterior~\cite{Planck_XIII:2016} restricts the expansion history to lie within the blue contours on the main panel of Fig.~\ref{fig:idl_h_z}, yielding the posterior on \hubble\ plotted in the left panel. The corresponding contours and posterior from \plancks\ \lcdm\ analysis are overlaid in grey. The inverse distance ladder \hubble\ constraint ($68.57 \pm 0.93$ \kmsmpc)\footnote{We find $q_0 = -0.50 \pm 0.08$ and $j_0 = 0.53 \pm 0.38$, consistent with both SH0ES (who fix $(q_0,j_0) =(-0.55, 1)$) and the Planck Collaboration (who find $q_0 = -0.54 \pm 0.02$ and fix $j_0 = 1$), and we thus discuss tension in terms of \hubble\ only.} is as precise as \plancks\ \lcdm\ constraint ($67.81 \pm 0.92$ \kmsmpc)---the flexibility of the model is offset by the extra data---and agrees to well within the 68\% credible intervals.

One potential concern here is that using the \planck\ \rdrag\ posterior introduces model inconsistency, as this assumes \lcdm\ and includes late-time information from lensing and the integrated Sachs-Wolfe effect. However, \rdrag\ is much less sensitive to late-time physics than \hubble: \rdrag\ constraints do not change significantly when the observational effects of late-time physical processes on the CMB are either removed or marginalized over~\cite{Verde_etal:2017}. For example, removing the lensing likelihood from the \planck\ \rdrag\ posterior shifts our \hubble\ posterior by less than 0.2-$\sigma$. We conclude that, to a good approximation, the \rdrag\ posterior employed here depends only on the assumption of standard pre-recombination physics.\footnote{Changes to the pre-recombination Universe do, however, have an impact: for example, allowing $N_{\rm eff}$ to vary yields a significantly broader posterior: $\hubble = 69.0 \pm 1.6$ \kmsmpc.} Using the WMAP9 \rdrag\ posterior~\cite{Hinshaw_etal:2013} in place of \plancks\ also does not change the conclusions, yielding $\hubble = 68.2 \pm 1.1$ \kmsmpc.

%%%%%%%%%%%%%%%%%%%%%%%%%%%%%%%%%%%%%%%%%%%%%%%%%%

{\bf Posterior Predictive Distribution.} With multiple discrepant \hubble\ estimates in hand, the task now is to define an intuitive measure of tension, which we will base on the PPD. The PPD is the sampling distribution for new data (${\mathbf d}^\prime$) given existing data (${\mathbf d}$) and a model ($I$)~\cite{Gelman_etal:2013} and so is given by averaging the likelihood of the new data over the posterior of the parameters ($\pars$) describing the model:
\begin{equation}
\prob({\mathbf d}^\prime|{\mathbf d}, I) = \int \prob({\mathbf d}^\prime|\pars, I) \, \prob(\pars|{\mathbf d}, I) \, d\pars.
\label{eq:ppd}
\end{equation}
With new data in hand, the PPD allows discrepancies between the data and model to be assessed: if the new data are not consistent with being drawn from the PPD, the model is not capable of fitting the data, and an alternative should be sought.

Although the PPD is typically employed to check the consistency of a replication of an experiment under the assumed model, there is no requirement for the two datasets to be derived from the same experimental process. Here, its utility in addressing tension between datasets becomes clear: given a dataset and a preferred model, the PPD can be used to simulate different measurements and hence assess whether the two datasets are consistent with being drawn from the same model.

\begin{figure}
\includegraphics[width=8.5cm]{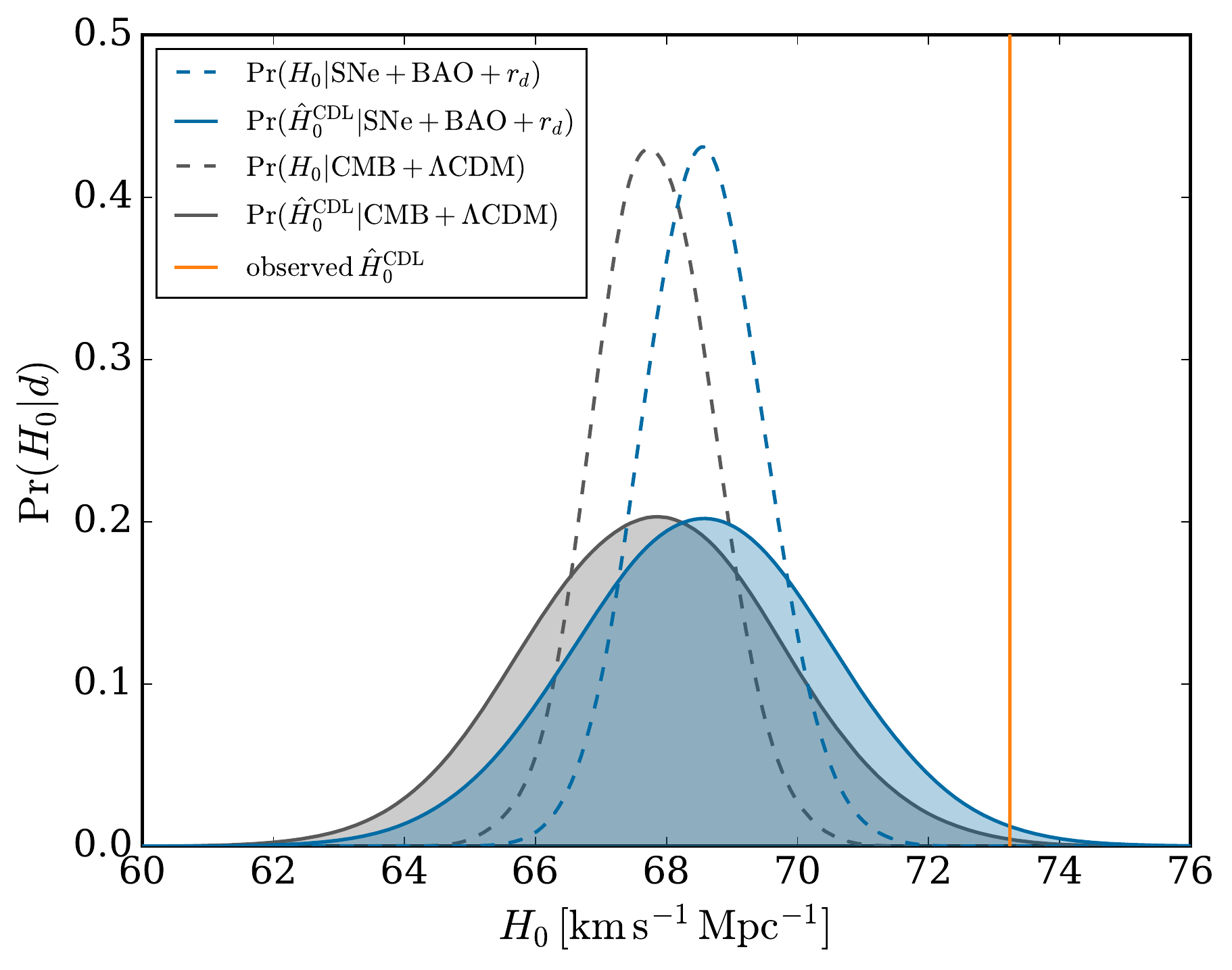}
\caption{PPDs (shaded) for the Cepheid distance ladder \hubbleobs, conditioned on inverse distance ladder data assuming a smooth expansion history (blue) or CMB data assuming \lcdm\ (grey). The SH0ES measurement is plotted as an orange solid line, and the \hubble\ posteriors from which the PPDs derive are plotted as dashed lines.}
\label{fig:idl_ppd}
\end{figure}

To demonstrate the PPD's utility, we use it to predict the SH0ES data given our inverse distance ladder data. For clarity, rather than predicting the full Cepheid distance ladder dataset\footnote{Predicting a less processed version of the dataset, e.g., the Cepheid magnitudes, could potentially yield greater insight into undiagnosed systematics at the cost of increasing the dimensionality of the PPDs.}, we predict the value of the resultant maximum likelihood estimate of the Hubble constant, $\hubbleobs^{\rm CDL}$. Converting the inverse distance ladder \hubble\ posterior into a PPD for $\hubbleobs^{\rm CDL}$, i.e., $\prob(\hubbleobs^{\rm CDL} | \msnobs, \abaoobs, I)$, is done by drawing one sample from the ``likelihood'' $\prob(\hubbleobs^{\rm CDL}|\hubble,I)$ for each sample from the inverse distance ladder posterior. Taking the likelihood to be a Gaussian with standard deviation 1.74 \kmsmpc, we obtain the PPD plotted in solid dark blue in Fig.~\ref{fig:idl_ppd}; the posterior from which it is derived is plotted in dashed dark blue. The actual $\hubbleobs^{\rm CDL}$ measured by SH0ES, overlaid as a solid orange line, is well into the tails of the PPDs: it is an unlikely draw from this sampling distribution.

In order to quantify the tension we calculate a simple statistic -- the ``PPD ratio'' -- defined as the ratio of the PPD at the observed $\hubbleobs^{\rm CDL}$ to its maximum:
\begin{equation}
\rho = \frac{\prob(\hubbleobs^{\rm CDL,obs} | \msnobs, \abaoobs, I)}{{\rm max}[\prob(\hubbleobs^{\rm CDL} | \msnobs, \abaoobs, I)]}.
\label{eq:rho}
\end{equation}
The PPD ratio has a number of advantages over other tension metrics~\cite[{e.g.,}][]{Marshall_etal:2006,Seehars_etal:2016,Riess_etal:2016,Feeney_etal:2017,Wang_etal:2017}: it can be generated at the cost of a single likelihood draw per posterior sample; it is simple to calculate even when the posterior is not convex or unimodal; and it is meaningful even in these general settings, unlike other summary statistics (e.g., $n$-$\sigma$ discrepancies or $p$-values). Finally, unlike model-comparison techniques, there is no need to specify an alternative model, nor is there strong dependence on the prior: informative data will make strong predictions even if the prior is improper.

The PPD ratio is also the likelihood ratio that would result from comparing a null model (that the SH0ES $\hubbleobs^{\rm CDL}$ is a random draw from its PPD) to an alternative ``just-so'' model in which the true $\hubble$ is fixed to the SH0ES value. As such, the PPD ratio can be interpreted as a lower bound on the posterior probability of the hypothesis that the two experiments measure the same \hubble\ without systematics. In this instance, the PPD ratio\footnote{Replacing the approximate Gaussian likelihood with the full asymmetric likelihood of Ref.~\cite{Feeney_etal:2017} reduces the quoted values of our PPD ratio less than 10\%.} is $1/17 \simeq 0.06$ at the SH0ES $\hubbleobs^{\rm CDL}$, so the probability that the distance ladders are unaffected by systematics, and that the apparent discrepancy is simply random, is at least 6\%. The PPD constructed from the \planck\ \lcdm\ posterior is shifted toward lower \hubble\ than the inverse distance ladder and so yields a lower ratio of 1/45. For comparison, the 3-$\sigma$ threshold commonly used in the Gaussian setting corresponds to a ratio of 1/90.

%%%%%%%%%%%%%%%%%%%%%%%%%%%%%%%%%%%%%%%%%%%%%%%%%%
%%%%%%%%%%%%%%%%%%%%%%%%%%%%%%%%%%%%%%%%%%%%%%%%%%

\section{Arbitrating Tension With Standard Sirens}
\label{sec:bns}

Observations of binary neutron star (BNS) mergers offer a method of measuring \hubble~\cite{Schutz:1986,Nissanke_etal:2010,Nissanke_etal:2013,Abbott_etal:2017a} that is completely independent of the Cepheid distance ladder and CMB. Fitting a merger's GW signal yields constraints on the luminosity distance ($d$) to the binary. Where a unique electromagnetic (EM) counterpart can be identified, a spectroscopic redshift for the host may be obtained, allowing a direct estimate of \hubble\ via
\begin{equation}
c \, z = \vpec + \hubble \, d.
\label{eq:gw_grb_hubble}
\end{equation}
The peculiar velocity (\vpec) can be left as a nuisance parameter~\cite{Guidorzi_etal:2017} or estimated~\cite{Abbott_etal:2017a} from ancillary data.

By simulating BNS data we can investigate the number of mergers needed to arbitrate the tension between the Cepheids and CMB using the PPD. Consider a set of $\nbns$ mergers with GW observations $\{\bm{x}\}$, peculiar velocity estimates $\{\vpecobs\}$ and perfectly observed redshifts $\{ \zobs \}$. Assuming Gaussian \vpecobs\ likelihoods (with uncertainties $\sigma_i$) and a Gaussian \vpec\ prior (of width $\sigma$), the marginal \hubble\ posterior becomes
\begin{multline}
\prob(\hubble |\{ \bm{x} \},  \{ \vpecobs \},  \{ \zobs \}, I) \propto \prob(\hubble|I) \, \prod_{i=1}^{\nbns} 
\label{eq:h_0_post} \\
\int dd_i \, \prob(d_i | \bm{x}_i, I) \, {\rm N} \! \left( \! \hubble \, d_i; c \, \zobs_i - \frac{\sigma^2 \, \vpecobs_i}{\sigma^2 + \sigma_i^2}, \frac{\sigma^2 \, \sigma_i^2}{\sigma^2 + \sigma_i^2} \right)
\end{multline}
(see Appendix~\ref{sec:appendix} for more detail) if the events are selected by their GW signal-to-noise ratio (SNR)~\cite[c.f.][]{Abbott_etal:2017a,Chen_etal:2017}. Converting this posterior into a PPD for the CMB or Cepheid distance ladder measurements is a straightforward integral with the relevant ``likelihood'' $\prob(\hubbleobs|\hubble,I)$.

We simulate a sample of BNS mergers and process it using the same Bayesian parameter-estimation pipeline as employed on real data, including the effects of amplitude and phase calibration uncertainties. We simulate BNS detections during the next three LIGO-Virgo (LV) observing runs assuming an underlying rate of \RBNS\ (consistent with the bounds from GW170817~\cite{TheLIGOScientific:2017qsa} at 90\% confidence), and a three-detector duty cycle of 40\%. Events are assumed to be independently distributed uniformly in comoving volume, with NS masses drawn from the Gaussian ${\rm N} (m_i \, ; 1.4\,{\rm M}_\odot,(0.2\,{\rm M}_\odot)^2)$ restricted to the range 1--3 ${\rm M}_\odot$. Binary orientations and NS spins are isotropically oriented, with spin magnitudes ${\leq} \, 0.05$~\cite{TheLIGOScientific:2017qsa}. Each simulated waveform is generated using a time-domain post-Newtonian approximation~\cite{Blanchet:2013haa,Buonanno:2009zt} and embedded in colored Gaussian noise realizations with power spectral densities~\cite[][Fig. 1]{Aasi:2013wya} appropriate to the detection date: ${\sim}$2019 (1 year); ${\sim}$2021 (1 year); and 2022+ (Design, 2 years). We deem BNS events ``GW detectable'' when two or more detectors have SNRs $\geq6$, and the network has SNR $\geq12$. This yields \NBNS\ detections. Fixing the sky position by assuming known host galaxies, we sample the parameter posteriors for each detection using a complete Bayesian MCMC analysis~\cite{Veitch:2014wba} with a frequency-domain post-Newtonian waveform model~\cite{Blanchet:2013haa,Buonanno:2009zt} spanning the range 30--2048 ${\rm Hz}$.\footnote{This takes a few CPU weeks per BNS posterior.} For estimating \hubble, we retain each event's distance posterior, marginalizing over all other parameters.

\begin{figure}
\includegraphics[width=8.5cm]{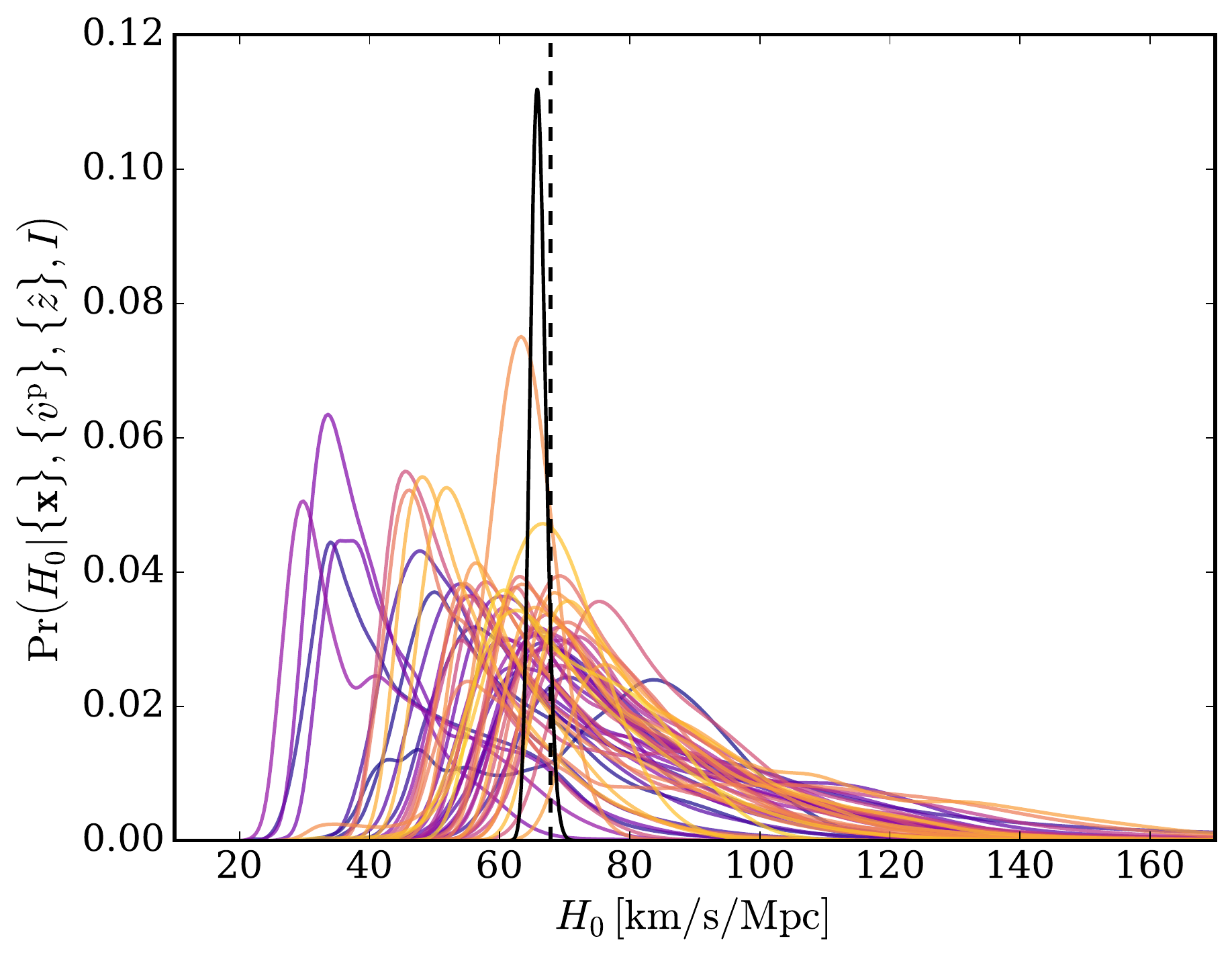}
\caption{\hubble\ posteriors for individual BNS mergers (purple to yellow, sorted by signal-to-noise) and the full sample (black solid; scaled by a factor of 1/3), assuming a true \hubble\ of 67.81 \kmsmpc\ (black dashed).}
\label{fig:bns_post}
\end{figure}

To complete the simulated dataset we need \vpecobs\ estimates and hence a true \hubble. For illustrative purposes, we use two true \hubble\ values, assuming either \planck\ or SH0ES is correct. We generate Gaussian measurement errors for each source's \vpecobs\ with standard deviation 200 ${\rm km\,s}^{-1}$~\cite{Carrick_etal:2015,Scolnic_etal:2017}. The \hubble\ posterior for the resulting simulated BNS dataset (assuming a true \hubble\ of 67.81 \kmsmpc) is plotted in Fig.~\ref{fig:bns_post}, along with posteriors for each individual event, colored by SNR. Our 1.8\% \hubble\ constraint from 51 mergers is in good agreement with the recent analysis of Ref.~\cite{Chen_etal:2017}. This complementary study uses an approximate 3D localization of GW sources~\cite{Chen_Holz:2016} to rapidly average over samples of mergers between compact objects of a single mass, with or without EM counterparts. Ref.~\cite{Chen_etal:2017} finds that $\sim$60 mergers between 1.4 ${\rm M}_\odot$ BNSs will, on average, constrain \hubble\ to 2\% assuming unique EM counterparts can be identified.

\begin{figure}
\includegraphics[width=8.5cm]{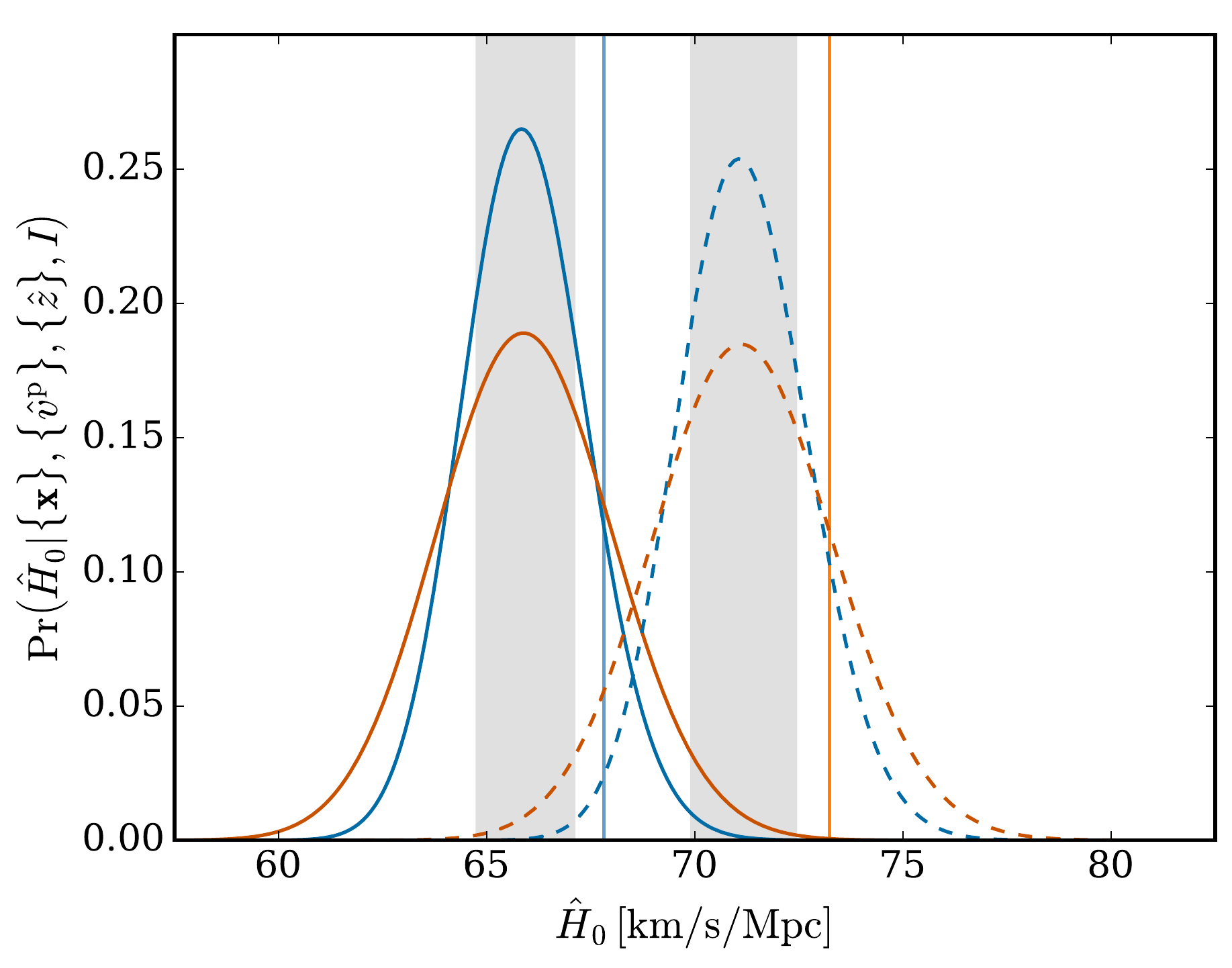}
\caption{PPDs for the CMB (dark blue) and Cepheid distance ladder (dark orange) \hubbleobs\ measurements, given the simulated BNS data. Solid/dashed curves assume the true \hubble\ to be the \planck/SH0ES measured value, indicated by the light blue/orange solid line. The 1-$\sigma$ variations in PPD means due to sample variance are shaded grey.}
\label{fig:bns_ppds}
\end{figure}

To convert the BNS \hubble\ posteriors to PPDs for the CMB and Cepheid distance ladder measurements, we take Gaussian likelihoods $\prob(\hubbleobs|\hubble,I)$ with standard deviations of 0.92 and 1.74 \kmsmpc, respectively. The results are plotted in Fig.~\ref{fig:bns_ppds}. The solid curves, for which we assume the \planck\ \hubble\ is correct, demonstrate the ability of this BNS sample to arbitrate the tension. The observed SH0ES \hubbleobs\ (solid light orange) would be an extremely unlikely draw from its sampling distribution (solid dark orange): the PPD ratio is $\sim$1/300, much lower than the 3-$\sigma$ equivalent ratio of 1/90. The \planck\ observation (solid light blue) would, as expected, be consistent with its PPD (solid dark blue). The BNS and CMB observations would decisively favor the underlying value of \hubble\ used in the simulations.

The {\it dashed} curves in Fig.~\ref{fig:bns_ppds}, in which we assume the SH0ES \hubble\ is correct, demonstrate another important aspect of this analysis: sample variance due to the limited number of detectable events. The posterior for our simulated sample happens to be scattered to low \hubble.\footnote{We have verified in further work that the \hubble\ posterior is statistically unbiased in this setting~\cite{Mortlock_etal:2018}.} Even though the BNS data strongly constrain \hubble---the posterior uncertainty is 1.2 \kmsmpc, less than a quarter of the tension---sample variance means we could not arbitrate in favor of one dataset. Indeed, we should expect to see realization-dependent variations in the PPD means on the scale of the posterior standard deviation. We confirm this using 1000 bootstrapped resamples of our dataset, shading the range of PPD means in grey in Fig.~\ref{fig:bns_ppds}. As such, while samples of $\sim$50 BNS mergers are certainly sufficient to arbitrate the tension, realization noise plays a role in determining whether it is possible for a given dataset. If the SH0ES \hubble\ measurement is correct, samples of $\sim$80 events will arbitrate the tension even if the BNS \hubble\ posterior is shifted by 1-$\sigma$ towards the \planck\ estimate by realization noise. If \planck\ is correct, significantly larger samples ($\sim$3000) are needed, as the PPD width is dominated by the SH0ES likelihood, which is independent of the BNS sample size.

%%%%%%%%%%%%%%%%%%%%%%%%%%%%%%%%%%%%%%%%%%%%%%%%%%
%%%%%%%%%%%%%%%%%%%%%%%%%%%%%%%%%%%%%%%%%%%%%%%%%%

\section{Conclusions}

We have demonstrated how existing and upcoming datasets can arbitrate the tension between estimates of \hubble\ from the CMB and local distance ladder. Throughout, we adopt the minimal cosmological model: a smooth expansion history and standard pre-recombination physics. We find that the inverse distance ladder formed from BOSS BAO measurements and the Pantheon SN sample yields an \hubble\ posterior near-identical to \planck\ and inconsistent with the observed local distance ladder value. We quantify this tension using a model-testing framework based on the posterior predictive distribution, which relies only on the sampling distribution for one dataset conditional on another, finding that the probability that the two distance ladders measure \hubble\ without systematics is at least 6\%. We then demonstrate how a typical sample of $\sim$50 BNS standard sirens, detectable by the LIGO and Virgo experiments within a decade, can independently arbitrate this tension. 

%%%%%%%%%%%%%%%%%%%%%%%%%%%%%%%%%%%%%%%%%%%%%%%%%%
%%%%%%%%%%%%%%%%%%%%%%%%%%%%%%%%%%%%%%%%%%%%%%%%%%

\section{Acknowledgments}

We are very grateful for useful discussions with Lauren Anderson, George Efstathiou, Archisman Ghosh, David W. Hogg, Daniel Holz, Benjamin Joachimi and Benjamin Wandelt. HVP is grateful for the hospitality of the Centro de Ciencias de Benasque Pedro Pascual, where elements of this work were conceived. HVP was partially supported by the European Research Council (ERC) under the European Community's Seventh Framework Programme (FP7/2007-2013)/ERC grant agreement number 306478-CosmicDawn, and the research environment grant ``Gravitational Radiation and Electromagnetic Astrophysical Transients (GREAT)'' funded by the Swedish Research council (VR) under Dnr 2016-06012. ARW and SMN acknowledge the generous financial support of the Netherlands Organization for Scientific Research through the NWO VIDI No. 639.042.612 (PI:Nissanke) and NWO TOP Grants No. 62002444 (PI:Nissanke). DS is supported by NASA through Hubble Fellowship grant HST-HF2-51383.001 awarded by the Space Telescope Science Institute, which is operated by the Association of Universities for Research in Astronomy, Inc., for NASA, under contract NAS 5-26555. The Flatiron Institute is supported by the Simons Foundation. This work was partially enabled by funding from the UCL Cosmoparticle Initiative.

%%%%%%%%%%%%%%%%%%%%%%%%%%%%%%%%%%%%%%%%%%%%%%%%%%
%%%%%%%%%%%%%%%%%%%%%%%%%%%%%%%%%%%%%%%%%%%%%%%%%%

\appendix

\begin{widetext}
\section{Analytic Standard Siren Posterior Derivation}
\label{sec:appendix}

Given a population of $\nbns$ binary neutron star (BNS) mergers with gravitational wave (GW) observations $\{\bm{x}\}$, peculiar velocity estimates $\{\vpecobs\}$ and redshifts $\{ \zobs \}$, the posterior on the Hubble constant (\hubble) is
\begin{equation}
\prob(\hubble |\{ \bm{x} \},  \{ \vpecobs \},  \{ \zobs \}, I) \propto \prob(\hubble|I) \, \prod_{i=1}^{\nbns} \int d\vpec_i \int dd_i \, \prob(d_i | \bm{x}_i, I) \, \prob(\vpecobs_i|\vpec_i,I) \, \prob(\zobs_i|\hubble,d_i,\vpec_i,I) \, \prob(\vpec_i | I),
\label{eq:h_0_post_full}
\end{equation}
where we have already marginalized over all parameters $\{\pars\}$ describing each GW source's waveform besides their distances $\{d\}$, assuming each set of observations is independent. Evaluating this posterior as written is not trivial, as it requires the true distance and peculiar velocity of each source to be inferred and depends on distance posteriors estimated from MCMC samples. One can, however, analytically marginalize over the true peculiar velocities if the peculiar velocity priors and the peculiar velocity and redshift likelihoods are assumed to be Gaussian. Taking these distributions to be ${\rm N}(\vpec_i;0,\sigma_{\vpec_i}^2)$, ${\rm N}(\vpecobs_i;\vpec_i,\sigma_{\vpecobs_i}^2)$ and ${\rm N}(\zobs_i;[\vpec_i + \hubble\,d_i]/c,\sigma_{\zobs_i}^2)$ respectively, we find that
\begin{equation}
\prob(\hubble |\{ \bm{x} \},  \{ \vpecobs \},  \{ \zobs \}, I) \propto \prob(\hubble|I) \, \prod_{i=1}^{\nbns} \int dd_i \, \prob(d_i | \bm{x}_i, I) \, {\rm N} \left( \hubble \, d_i; c \, \zobs_i - \frac{\sigma_{\vpec_i}^2 \, \vpecobs_i}{\sigma_{\vpec_i}^2 + \sigma_{\vpecobs_i}^2}, c^2 \sigma_{\zobs_i}^2 + \frac{\sigma_{\vpec_i}^2 \, \sigma_{\vpecobs_i}^2}{\sigma_{\vpec_i}^2 + \sigma_{\vpecobs_i}^2} \right).
\label{eq:h_0_post_gauss_v_z}
\end{equation}

If the distance posteriors from BNS mergers were typically Gaussian, i.e. $\prob(d_i | \bm{x}_i, I) = {\rm N}(d_i;\dobs_i,\sigma_{\dobs_i}^2)$, Eq.~\ref{eq:h_0_post_gauss_v_z} would resolve into the particularly simple form
\begin{equation}
\prob(\hubble |\{ \bm{x} \},  \{ \vpecobs \},  \{ \zobs \}, I) \propto \prob(\hubble|I) \, \prod_{i=1}^{\nbns} {\rm N} \left( \hubble \, \dobs_i; c \, \zobs_i - \frac{\sigma_{\vpec_i}^2 \, \vpecobs_i}{\sigma_{\vpec_i}^2 + \sigma_{\vpecobs_i}^2}, \hubble^2 \sigma_{\dobs_i}^2 + c^2 \sigma_{\zobs_i}^2 + \frac{\sigma_{\vpec_i}^2 \, \sigma_{\vpecobs_i}^2}{\sigma_{\vpec_i}^2 + \sigma_{\vpecobs_i}^2} \right).
\label{eq:h_0_post_gauss_v_z_d}
\end{equation}
This is unfortunately not the case, as the distance posteriors are typically highly non-Gaussian. Nevertheless, this expression remains useful as it provides an estimate of the uncertainty on \hubble\ expected from a sample of $n$ mergers. Each independent event constrains \hubble\ with variance (c.f. Ref.~\cite{Chen_etal:2017}, Eq. 1)
\begin{equation}
\sigma_{\hubble}^2 \simeq \frac{ \hubble^2 \sigma_{\dobs_i}^2 }{ \dobs_i^2 } + \frac{ c^2 \sigma_{\zobs_i}^2 }{ \dobs_i^2 } + \frac{\sigma_{\vpec_i}^2 \, \sigma_{\vpecobs_i}^2}{\dobs_i^2 (\sigma_{\vpec_i}^2 + \sigma_{\vpecobs_i}^2)}.
\label{eq:h_0_var_single_gauss_v_z_d}
\end{equation}
Neglecting the slight skewness introduced by the presence of \hubble\ in the denominator of Eq.~\ref{eq:h_0_post_gauss_v_z_d}, a sample of $n$ events with characteristic observed distances $\dobs$, redshifts $\zobs$ and peculiar velocities $\vpecobs$ will therefore yield a combined constraint of
\begin{equation}
\frac{\sigma_{\hubble}^2}{\hubble^2} \simeq \frac{1}{n} \left( \frac{ \sigma_{\dobs}^2 }{ \dobs_i^2 } + \frac{ 1 }{ \zobs^2 } \left[ \sigma_{\zobs}^2 + \frac{\sigma_{\vpec}^2 \, \sigma_{\vpecobs}^2}{c^2  (\sigma_{\vpec}^2 + \sigma_{\vpecobs}^2)} \right] \right).
\label{eq:h_0_var_gauss_v_z_d}
\end{equation}

In order to process generic BNS distance posteriors analytically, in this work we fit the distance posteriors using Gaussian kernel density estimates (KDEs), where the KDE for the $i^{\rm th}$ GW source's distance posterior is characterized by its number of kernels, $\nk_i$, bandwidth, $b_i$, and each kernel's index, $j$, and mean, $\mu_{ij}$. Eq.~\ref{eq:h_0_post_gauss_v_z} then simplifies to
\begin{equation}
\prob(\hubble |\{ \bm{x} \},  \{ \vpecobs \},  \{ \zobs \}, I) \propto \prob(\hubble|I) \, \prod_{i=1}^{\nbns} \frac{1}{\nk_i} \sum_{j=1}^{\nk_i} {\rm N} \left( \hubble \, \mu_{ij}; c \, \zobs_i - \frac{\sigma_{\vpec_i}^2 \, \vpecobs_i}{\sigma_{\vpec_i}^2 + \sigma_{\vpecobs_i}^2}, \hubble^2 b_i^2 + c^2 \sigma_{\zobs_i}^2 + \frac{\sigma_{\vpec_i}^2 \, \sigma_{\vpecobs_i}^2}{\sigma_{\vpec_i}^2 + \sigma_{\vpecobs_i}^2} \right).
\label{eq:h_0_post_super_gauss_inc_z_kde}
\end{equation}
We assume that the EM counterparts have spectroscopic redshift measurements, setting $\sigma_{\zobs_i} = 0$ (equivalent to using delta-function redshift likelihoods, $\prob(\zobs_i|\hubble,d_i,\vpec_i,I) = \delta(\zobs_i - [\vpec_i + \hubble d_i]/c)$, throughout). In the limit that the peculiar velocity likelihoods are much narrower than the priors ($\sigma_{\vpecobs_i} \ll \sigma_{\vpec_i}$, i.e., we have informative peculiar velocity measurements) the KDE-approximated \hubble\ posterior becomes
\begin{equation}
\prob(\hubble |\{ \bm{x} \},  \{ \vpecobs \},  \{ \zobs \}, I) \propto \prob(\hubble|I) \, \prod_{i=1}^{\nbns} \frac{1}{\nk_i} \sum_{j=1}^{\nk_i} {\rm N} \left( \hubble \, \mu_{ij}; c \, \zobs_i - \vpecobs_i, \hubble^2 \, b_i^2 + \sigma_{\vpecobs_i}^2 \right).
\label{eq:h_0_post}
\end{equation}
If, instead, no peculiar velocity observations are available ($\sigma_{\vpecobs_i} = \infty$), the approximate posterior is
\begin{equation}
\prob(\hubble |\{ \bm{x} \}, \{ \zobs \}, I) \propto \prob(\hubble|I) \, \prod_{i=1}^{\nbns} \frac{1}{\nk_i} \sum_{j=1}^{\nk_i} {\rm N} \left( \hubble \, \mu_{ij}; c \, \zobs_i, \hubble^2 \, b_i^2 + \sigma_{\vpec_i}^2 \right).
\label{eq:h_0_post_no_v_pec_obs}
\end{equation}
\end{widetext}

%%%%%%%%%%%%%%%%%%%%%%%%%%%%%%%%%%%%%%%%%%%%%%%%%%
%%%%%%%%%%%%%%%%%%%%%%%%%%%%%%%%%%%%%%%%%%%%%%%%%%

\bibliographystyle{apsrev4-1}
\bibliography{h_0_ppd}

\end{document}